\documentclass[letterpaper,times]{IONconf}

\usepackage[pdftex]{graphicx}
\DeclareGraphicsExtensions{.pdf,.jpeg,.png}

\usepackage{amsmath}
\usepackage{amssymb}
\usepackage[ruled]{algorithm2e}
\usepackage{algpseudocode}
\usepackage{array}
\usepackage{textcomp}
\usepackage{makecell}
\usepackage{hhline}
\usepackage{multicol}
\usepackage{multirow}
\usepackage{booktabs}
\usepackage{threeparttable}
\usepackage{verbatim}
\usepackage{xcolor}
\usepackage{pifont}

\newcolumntype{P}[1]{>{\centering\arraybackslash}p{#1}}

\newcommand{\inv}[1]{#1^{-1}}
\newcommand{\trp}[1]{{#1}^{\mathsf{T}}}

\newcommand{\thickhline}{
    \noalign {\ifnum 0=`}\fi \hrule height 1pt
    \futurelet \reserved@a \@xhline
}
\DeclareMathOperator*{\argmax}{arg\,max}
\DeclareMathOperator*{\argmin}{arg\,min}

\makeatletter
\def\env@dmatrix{\hskip -\arraycolsep
  \let\@ifnextchar\new@ifnextchar
  \extrarowheight=2ex
  \array{*\c@MaxMatrixCols{>{\displaystyle}c}}}

\newenvironment{bdmatrix}
  {\left[\env@dmatrix}
  {\endmatrix\right]}
\makeatother

\DeclareCaptionLabelSeparator{custom}{ }
\DeclareCaptionFormat{custom}
{%
    \textbf{#1#2}\textit{\small #3}
}
\usepackage{url}
\usepackage{natbib}
\usepackage[hidelinks]{hyperref}

\title{Performance Evaluation and Hybrid Application of the Greedy and Predictive UAV Trajectory Optimization Methods for Localizing a Target Mobile Device}

\author{
    Halim~Lee and Jiwon~Seo, \textit{Yonsei~University}%
    }

\begin{document}

\maketitle

\section*{biography}

\biography{Halim Lee}{is a Ph.D. candidate in the School of Integrated Technology, Yonsei University, Incheon, Korea. She received the B.S. degree in Integrated Technology from Yonsei University. She was a visiting graduate student at the University of California, Irvine in 2020. Her research interests include urban navigation, navigation safety, and localization and tracking.}

\biography{Jiwon Seo}{is an associate professor with the School of Integrated Technology, Yonsei University, Incheon, Korea. He received the B.S. degree in mechanical engineering (division of aerospace engineering) in 2002 from Korea Advanced Institute of Science and Technology, Daejeon, Korea, and the M.S. degree in aeronautics and astronautics in 2004, the M.S. degree in electrical engineering in 2008, and the Ph.D. degree in aeronautics and astronautics in 2010 from Stanford University, Stanford, CA, USA.
His research interests include GNSS and complementary PNT systems. Prof. Seo is a member of the International Advisory Council of the Resilient Navigation and Timing Foundation, Alexandria, VA, USA, and a member of several advisory committees of the Ministry of Oceans and Fisheries and the Ministry of Land, Infrastructure and Transport, Korea.}

\section*{Abstract}

This study investigates unmanned aerial vehicle (UAV) trajectory planning strategies for localizing a target mobile device in emergency situations. 
The global navigation satellite system (GNSS)-based accurate position information of a target mobile device in an emergency may not be always available to first responders.
For example, 1) GNSS positioning accuracy may be degraded in harsh signal environments and 2) in countries where emergency positioning service is not mandatory, some mobile devices may not report their locations. 
Under the cases mentioned above, one way to find the target mobile device is to use UAVs. 
Dispatched UAVs may search the target directly on the emergency site by measuring the strength of the signal (e.g., LTE wireless communication signal) from the target mobile device. 
To accurately localize the target mobile device in the shortest time possible, UAVs should fly in the most efficient way possible. 
The two popular trajectory optimization strategies of UAVs are greedy and predictive approaches. 
However, the research on localization performances of the two approaches has been evaluated only under favorable settings (i.e., under good UAV geometries and small received signal strength (RSS) errors); more realistic scenarios still remain unexplored.
In this study, we compare the localization performance of the greedy and predictive approaches under realistic RSS errors (i.e., up to 6 dB according to the ITU-R channel model). 
From the simulation result, the greedy approach performs better in reducing the localization error at the initial stage of the search; however, the predictive approach performs better once the localization error converges to a certain value. 
Based on these observations, we propose a hybrid application involving both the approaches. 
The performance of the proposed hybrid approach was evaluated under less diverse UAV geometries and realistic RSS errors.
During simulation tests, the hybrid approach showed localization accuracy improvements of 30.8\% and 55.0\% over the greedy-only and predictive-only approaches, respectively.

\section{INTRODUCTION}

The exact and fast localization of a target mobile device is critical in emergency response. In the United States, the Federal Communications Commission (FCC) currently requires commercial mobile radio service (CMRS) providers to satisfy the specified accuracy requirements (e.g., vertical accuracy of 3 m and horizontal accuracy of 50 m for 80\% of all indoor 911 calls) on mobile devices \citep{FCC19}. To meet those requirements and provide various location-based services (LBS), most modern mobile devices contain global navigation satellite system (GNSS) chipsets. However, the accuracy and availability of GNSS may be largely degraded in harsh environments, such as urban or indoors, owing to signal blockage and attenuation by buildings \citep{Soloviev10:Tight, Strandjord20:Improved, Lee22:Urban, Kim21:GPS,  Lee22:Nonlinear}. Further, in countries where emergency positioning services are non-mandatory for manufacturers or carriers, the position information of a user in an emergency may not be accessible on some devices.

Therefore, to find the target mobile device in GNSS-denied conditions, unmanned aerial vehicle (UAV)-based target localization \citep{Dogancay12:UAV, Shahidian16:Path, Wang18:Performance, Uluskan20:Noncausal, Koohifar16:Receding, Lee22:Evaluation} is considered in this study. We assume UAVs \citep{Kim20:Motion, Kim14:Multi} are dispatched to the emergency location to localize the target by measuring the target’s signal. Among several signal measurements (e.g., time-of-arrival (TOA) \citep{Shamaei18:LTE, Wang18:Performance, Uluskan20:Noncausal, Lee20:Preliminary, Lee20:Neural}, time-difference-of-arrival (TDOA) \citep{Son18:Novel, Son19:Universal, Kim20:Development, Park20:Effect}, angle-of-arrival (AOA) \citep{Park21:Single, Park18:Dual}, and received signal strength (RSS) \citep{Wang18:Performance, Uluskan20:Noncausal, Koohifar16:Receding, Kang20:Practical, Lee22:Performance}), this study considers the RSS measurement because of its hardware and software design simplicity.

The flying trajectory of UAVs should be optimized in a manner to localize the target as precisely as possible within the given time frame. There are numerous well-established trajectory optimization strategies \citep{Li20:Path, Liu18:A, Aggarwal20:Path, Rutkowski12:The, Rutkowski15:Path, Lee20:Integrity}. In this study, we considered the Cramer-Rao lower bound (CRLB), which is a lower bound on the variance of position estimates of an unbiased estimator, as a metric for optimization \citep{Dogancay12:UAV, Shahidian16:Path, Wang18:Performance, Uluskan20:Noncausal, Koohifar16:Receding}.

\begin{figure}
    \centering
    \includegraphics[width=0.8\linewidth]{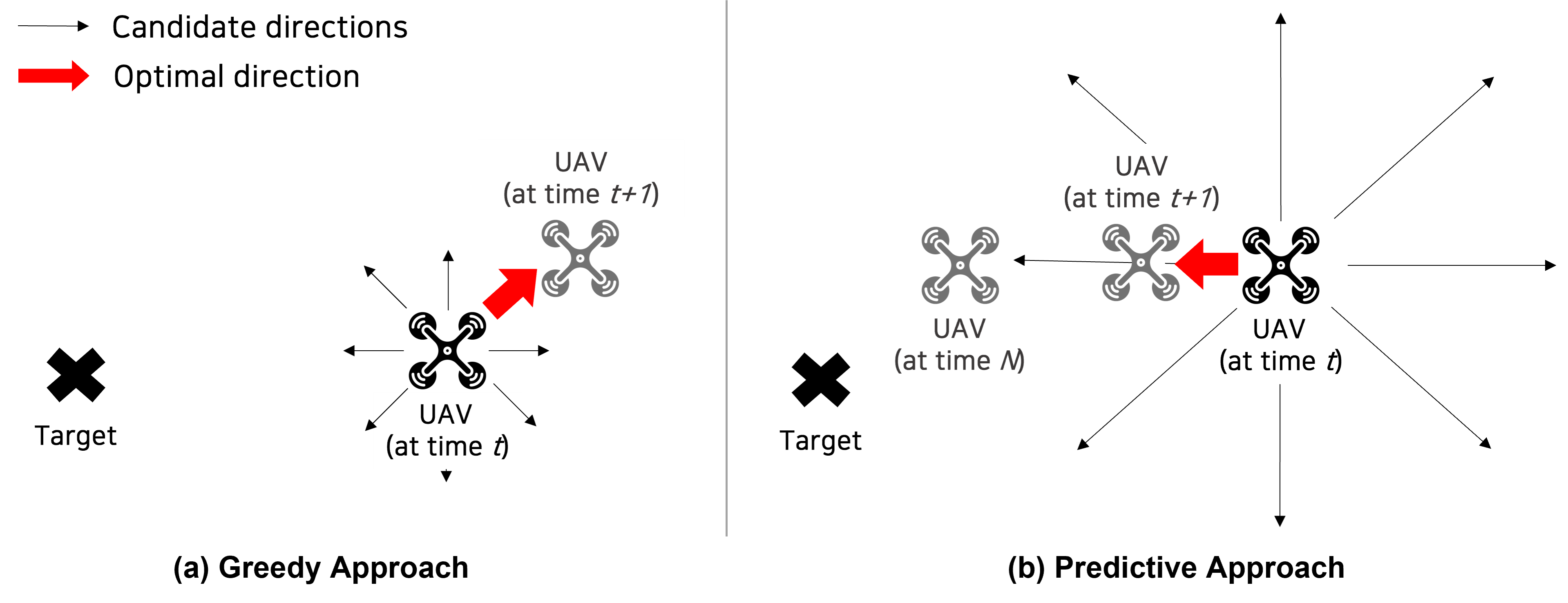}
    \caption{Conceptual Difference Between (a) Greedy and (b) Predictive Approaches}
    \label{fig:greedy_and_predictive}
\end{figure}

Using CRLB, two UAV trajectory optimization schemes were considered: greedy \citep{Dogancay12:UAV, Shahidian16:Path, Wang18:Performance} and predictive (i.e., noncausal \citep{Uluskan20:Noncausal} or receding horizon \citep{Koohifar16:Receding}) approaches. Figure \ref{fig:greedy_and_predictive} shows the conceptual difference between the two approaches. Both approaches optimize the UAV trajectories in the direction of minimizing the CRLB. However, the predictive approach optimizes the next waypoint ($t+1$ in Figure \ref{fig:greedy_and_predictive}(b)) based on the predicted CRLB of the final epoch (i.e., CRLB at the maximum allowable search time, which is $N$ in Figure \ref{fig:greedy_and_predictive}(b)), whereas the greedy approach considers only the CRLB of the next epoch ($t+1$ in Figure \ref{fig:greedy_and_predictive}(a)) when calculating the next waypoint.

In \citet{Uluskan20:Noncausal} and \citet{Koohifar16:Receding}, it was shown that the predictive approach has better localization accuracy compared to the greedy approach. However, these simulations were performed under less realistic settings; diverse UAV geometries and small RSS errors were assumed \citep{Wang18:Performance, Uluskan20:Noncausal, Koohifar16:Receding}. 
Table \ref{tab:RSS_error} compares the RSS error assumption of the previous studies \citep{Wang18:Performance, Uluskan20:Noncausal, Koohifar16:Receding} with the realistic line-of-sight (LOS) scenarios recommended by the International Telecommunication Union Radiocommunication Sector (ITU-R) channel model \citep{ITUR09}. 
The RSS error in dB refers to the standard deviation of the ratios in dB of the measured RSS values to the expected RSS values. The expected RSS value is the ideal RSS value obtained by the log-distance path loss model. 
The RSS error is assumed to be normally distributed in the dB scale.
The variable $d$ in Table \ref{tab:RSS_error} is the distance between the target and the UAV. 

Most studies have considered an RSS error of 0.01 dB. However, this assumption is more optimistic compared to the realistic channel models (note that the ITU-R channel model \citep{ITUR09} recommends up to 6 dB RSS error in suburban or rural macro scenarios). 
Even in scenarios that considered a realistic RSS error (e.g., 5 dB) \citep{Uluskan20:Noncausal}, the UAV geometry was assumed favorable (e.g., eight UAVs started the mission by surrounding the target). 

\begin{table}
\small
\centering
\caption{Comparison of RSS Error Assumptions in the Literature}\label{tab:RSS_error}
\begin{center}{
\renewcommand{\arraystretch}{1.4}
 \begin{tabular}{ c | P{1.3cm} P{1.3cm} P{1.6cm} P{1.5cm} | P{1.8cm} P{1.5cm} P{2.3cm}}
 \hhline{========}
 \rule{0pt}{15pt} {} & \multicolumn{4}{c|}{\thead{ITU-R LOS Scenarios \citep{ITUR09}}} & {} & {} & {} \\ 
 \multirow{-2}{*}{\thead{Reference}} & \thead{Urban \\ Micro} & \thead{Urban \\ Macro} & \thead{Suburban \\ Macro} & \thead{Rural \\ Macro} & \multirow{-2}{*}{\thead{\citeauthor{Wang18:Performance} \\ (\citeyear{Wang18:Performance})}} & \multirow{-2}{*}{\thead{\citeauthor{Uluskan20:Noncausal} \\ (\citeyear{Uluskan20:Noncausal})}} & \multirow{-2}{*}{\thead{\citeauthor{Koohifar16:Receding} \\ (\citeyear{Koohifar16:Receding})}} \\
 \hline
 \rule{0pt}{20pt} \thead{RSS Error \\ (dB)} & 3 & 4 & 4--6 & 4--6 & $0.01 \times d$ & 0.01 or 5 & 0.01\\
 \hhline{========}
\end{tabular}}
\begin{tablenotes}
\small
\item \quad \textit{Note}: The variable $d$ is the distance between the UAV and target.
\end{tablenotes}
\end{center}
\end{table}

We evaluated the localization performance of greedy and predictive approaches in scenarios with realistic RSS error assumptions. 
According to simulation results, the greedy approach showed better localization accuracy at the initial stage of the search. 
However, after the localization error converged to a certain value, the predictive approach showed better accuracy than the greedy approach.
Hence, we propose a hybrid scheme of combined greedy and predictive approaches.

The remainder of this paper describes the following: Section \ref{sec:problem} describes the UAV-based target localization problem. Section \ref{sec:CRLB} focuses on the calculation of CRLB, and Section \ref{sec:optimization} introduces greedy and predictive approaches. Section \ref{sec:evaluation} evaluates the performance of greedy and predictive approaches in various scenarios and proposes a hybrid approach. Finally, Section \ref{sec:conclusion} concludes this paper.

\section{Problem Description} 
\label{sec:problem}

We assumed the following emergency localization situation. $M$ UAVs are equipped with receivers capable of measuring RSS of a signal transmitted from the target. It is assumed that UAVs can estimate their own positions through their navigation systems (e.g., GNSS \citep{Enge1994:global, Sun21:Markov, Park22:Horizontal, Seo14:Future, Lee22:Optimal, Park21:Single, Kwak18:Autonomous, Kim21:GPS} or alternative navigation systems \citep{Zhang11:Novel, Kim14:Multi, Lee19:Safety, Huang19:Deep, Jia21:Ground, Strandjord21:Evaluating, Rhee21:Enhanced, Jeong20:RSS, Kim22:First, Lee22:SFOL}). 
When the maximum allowable search time is limited to $N$ seconds, the position of the $m$-th UAV at time $t \in \left[ 0, N \right]$ is expressed as $\mathbf{x}^\mathrm{UAV}_{m, t} = \trp{ \left[ x^\mathrm{UAV}_{m, t}, y^\mathrm{UAV}_{m, t} \right] }$. 
Note, although two-dimensional target localization is assumed in this study, our method can simply be expanded to three dimensions by adding a z-axis variable. 
When the actual position of the target is $\mathbf{r} = \trp{ \left[ x_\mathrm{r}, y_\mathrm{r} \right] }$, the RSS measured by the $m$-th UAV at time $t$ (i.e., $\hat{P}_{m, t}$), can be modeled as:
\begin{equation} \label{eq:Path_loss_model}
\begin{split}
\hat{P}_{m, t} &= P_{m,t} + n_{m, t} = P_{0} - 10 \beta \log_{10} \frac{d_{m, t}}{d_0} + n_{m, t},\\
d_{m, t} &= \|\mathbf{x}^\mathrm{UAV}_{m,t}-\mathbf{r}\|, \\
n_{m, t} &\sim \mathcal{N}(0, \sigma_\mathrm{dB}^2),
\end{split}
\end{equation}
where $P_{0}$ (in dBm) is the RSS at the distance $d_0$ from the target (in this study, $d_0$ is assumed to be 1 m without loss of generality); $\beta$ is the path loss exponent; $n_{m, t}$ is the log-normal shadowing that can be modeled as normal distribution in the dB scale with standard deviation of $\sigma_\mathrm{dB}$; and $\|\cdot\|$ denotes Euclidean norm.

Our goal is to determine an optimal UAV position at time $t+1$ (i.e., $\mathbf{x}^\mathrm{UAV}_{m, t+1}$) in the direction of lowering the target localization error the most using the RSS measurements up to time $t$.
The geometry between the target and the UAVs has a significant impact on RSS-based localization performance. The effects of geometry are reflected in the CRLB, which is the lower bound on any unbiased estimator's error variance.

\section{Cramer-Rao Lower Bound for RSS Localization} \label{sec:CRLB}

The CRLB is the inverse of the Fisher information matrix (FIM). The Fisher information quantifies how much information an observed random variable $\hat{P}_{m, t}$ has about an unknown parameter $\mathbf{r}$ \citep{Wang18:Performance, Uluskan20:Noncausal, Ly17:A}. Our objective is to optimize the UAVs' trajectories in the direction of increasing Fisher information (i.e., reducing CRLB) to accurately estimate the target's position. Considering any unbiased estimate of $\mathbf{r}$ (i.e., $\hat{\mathbf{r}}$), the covariance of $\hat{\mathbf{r}}$ is bounded by the inverse of the FIM \citep{Wang18:Performance, Uluskan20:Noncausal}:
\begin{equation} \label{eq:CRLB}
\begin{split}
\mathbb{E}\left[ \left( \hat{\mathbf{r}} - \mathbf{r} \right) \trp{\left( \hat{\mathbf{r}} - \mathbf{r} \right)} \right] \geq \mathbf{J}_{t}\left(\mathbf{r}\right)^{-1},
\end{split}
\end{equation}
where the $(i,j)$-th element of the FIM $\mathbf{J}_{t}$ at time $t$ is given by \citep{Ly17:A, Wang18:Performance}:
\begin{equation} \label{eq:FIM}
\begin{split}
\left[ \mathbf{J}_{t}\left(\mathbf{r}\right) \right]_{i,j} = \mathbb{E} \left[ \left( \frac{\partial}{\partial r_i} \ln \left( f \left( \hat{\mathbf{P}}_{t}{;} \mathbf{r}  \right) \right) \right) \left( \frac{\partial}{\partial r_j} \ln \left( f \left( \hat{\mathbf{P}}_{t}{;} \mathbf{r}  \right) \right) \right) \bigg \rvert \: \mathbf{r} \right],
\end{split}
\end{equation}
where $\hat{\mathbf{P}}_{t} = \trp{\left[ \hat{P}_{1,t}, \cdots, \hat{P}_{M,t} \right]}$ is the set of measurements collected by $M$ UAVs at time $t$; and $f \left( \hat{\mathbf{P}}_{t}{;} \mathbf{r}  \right)$ is the probability density function (PDF) of $\hat{\mathbf{P}}_{t}$ conditioned on the value of $\mathbf{r}$, which is a multivariate normal distribution as follows \citep{Ly17:A}:
\begin{equation} \label{eq:PDF}
\begin{split}
f \left( \hat{\mathbf{P}}_{t}{;} \mathbf{r}  \right) = \frac{1}{ (2\pi)^{M/2} \cdot \sigma_\mathrm{dB}} \exp \left[ {- \frac{1}{2} \trp{\left( \hat{\mathbf{P}}_t - \mathbf{P}_t \right)} \inv{\sigma_\mathrm{dB}} \left( \hat{\mathbf{P}}_t - \mathbf{P}_t \right) } \right],
\end{split}
\end{equation}
where $\mathbf{P}_{t} = \trp{\left[ P_{1,t}, \cdots, P_{M,t} \right]}$.

By combining Equations (\ref{eq:Path_loss_model}) and (\ref{eq:FIM}), the FIM for estimating the parameter $\mathbf{r}$ using RSS measurements $\hat{\mathbf{P}}_{t}$ can be reformulated as follows \citep{Uluskan20:Noncausal}:
\begin{equation} \label{eq:FIM_matrixform}
\begin{split}
\mathbf{J}_{t}\left(\mathbf{r}\right) &= K 
\begin{bdmatrix}
\sum_{m=1}^{M} \frac{{\left(x^\mathrm{UAV}_{m, t}-x_\mathrm{r}\right)}^2}{{d_{m,t}}^4} & \sum_{m=1}^{M} \frac{{\left(x^\mathrm{UAV}_{m, t}-x_\mathrm{r}\right)}\cdot{\left(y^\mathrm{UAV}_{m, t}-y_\mathrm{r}\right)}}{{d_{m,t}}^4} \\[2em]
\sum_{m=1}^{M} \frac{{\left(x^\mathrm{UAV}_{m, t}-x_\mathrm{r}\right)}\cdot{\left(y^\mathrm{UAV}_{m, t}-y_\mathrm{r}\right)}}{{d_{m,t}}^4} & \sum_{m=1}^{M} \frac{{\left(y^\mathrm{UAV}_{m, t}-y_\mathrm{r}\right)}^2}{{d_{m,t}}^4}
\end{bdmatrix}
, \\[0.8em]
K &= \frac{1}{\sigma_\mathrm{dB}} \left( \frac{10 \beta}{\ln\left( 10 \right)} \right).
\end{split}
\end{equation}

To calculate the FIM of Equation (\ref{eq:FIM_matrixform}), the true position of the target, $\mathbf{r}$, must be known to UAVs. However, knowing $\mathbf{r}$ in a real target localization situation is impossible. 
Following the approach of previous studies \citep{Uluskan20:Noncausal, Wang18:Performance, Koohifar16:Receding}, we calculated FIM using the estimate of $\mathbf{r}$ at time $t$ (i.e., $\mathbf{\hat{r}}_t$) instead of the true $\mathbf{r}$. 
We updated $\mathbf{\hat{r}}_t$ every epoch by maximum likelihood estimation (MLE) as follows \citep{Lee22:Evaluation}:
\begin{equation} \label{eq:MLE}
\begin{split}
    \mathbf{\hat{r}}_t &= \argmin_{\mathbf{r}} \sum_{i=0}^{t} \sum_{m=1}^{M} n_{m,t}^2 \\
    &= \argmin_{\mathbf{r}} \sum_{i=0}^{t} \sum_{m=1}^{M} \left( P_{m,i} - P_{0} + 10 \beta \log_{10} \frac{d_{m, t}}{d_0} \right) ^2.
\end{split}
\end{equation}
MLE in Equation (\ref{eq:MLE}) can be solved using numerical methods \citep{Baldi95:Gradient, Garg12:An, Lee22:Evaluation}. In this study, we implemented a grid search-based MLE solver.

As optimization metrics from the FIM, some optimal designs (e.g., A-, E-, and D-optimality) can be used. In previous studies \citep{Uluskan20:Noncausal, Wang18:Performance, Koohifar16:Receding}, the D-optimality for UAV trajectory optimization was commonly used. The D-optimality minimizes the volume of the uncertainty ellipsoid of the target's position estimate. To minimize the volume of the uncertainty ellipsoid, the determinant of $\mathbf{J}_{t}\left(\mathbf{r}\right)$ should be maximized.

\section{Greedy and Predictive Trajectory Optimization Approaches} \label{sec:optimization}

This section introduces two well-known trajectory optimization approaches: greedy and predictive approaches. 
As the number of RSS measurements increases over time, the volume of uncertainty ellipsoid of the position of target gradually decreases (or equivalently, the determinant of $\mathbf{J}_{t}\left(\mathbf{r}\right)$ gradually increases). 
In emergency situations, diminishing the volume of the uncertainty ellipsoid rapidly becomes important. 
To achieve this, the greedy and predictive approaches formulate optimization problems in determining the next moving directions of UAVs.

\subsection{Greedy Approach}

Given the previous trajectories and the current RSS measurements, the greedy approach optimizes the best possible moving direction $\alpha_{t+1}$ of each UAV, which can be formulated as \citep{Uluskan20:Noncausal, Wang18:Performance}:

\begin{equation} \label{eq:Greedy}
\begin{split}
\alpha_{t+1} = \underset{\alpha}{\argmax} \left[ \mathrm{det} \left( \mathbf{J}_{t+1}\left(\hat{\mathbf{r}}, \alpha\right) +  \sum_{i=0}^{t} \mathbf{J}_{i}\left(\hat{\mathbf{r}}\right) \right)  \right],
\end{split}
\end{equation}
where $\mathbf{J}_{t+1}\left(\hat{\mathbf{r}}, \alpha\right)$ is the FIM at the next epoch $t+1$ and a function of $\alpha$; and the position of the \textit{m}-th UAV at next epoch $t+1$ is given by \citep{Uluskan20:Noncausal}:
\begin{equation}
\begin{split}
x^\mathrm{UAV}_{m,t+1} \left( \alpha \right) &= x^\mathrm{UAV}_{m,t} + l \cdot \mathrm{cos} \left( \alpha \right)\\[0.6em]
y^\mathrm{UAV}_{m,t+1} \left( \alpha \right) &= y^\mathrm{UAV}_{m,t} + l \cdot \mathrm{sin} \left( \alpha \right),
\end{split}
\end{equation}
where $l$ is the separation between two subsequent waypoints of the UAV.

The optimal solution of Equation (\ref{eq:Greedy}) can be obtained by numerical methods. 
After calculating the optimal moving direction of the UAV, the UAV moves in the corresponding direction and adopts a new RSS measurement. 
This procedure continues until time reaches the maximum allowable search time $N$. 
This optimization approach is called ``greedy'' because it only considers  the FIM after one epoch, which yields a ``short-sighted'' optimal choice.

\subsection{Predictive Approach}

In contrast to the greedy approach, the predictive approach seeks a ``long-sightedness'' by predicting the FIM at the end of search (i.e., the maximum allowable search time). The predictive approach for optimizing the best possible moving direction of each UAV can be formulated as \citep{Uluskan20:Noncausal}:
\begin{equation} \label{eq:Predictive}
\begin{split}
\alpha_{t+1} = \underset{\alpha}{\argmax} \left[ \, \mathrm{det} \left( \: \sum_{k=1}^{N-t} \mathbf{J}_{t+k}\left(\hat{\mathbf{r}}, \alpha\right) +  \sum_{i=0}^{t} \mathbf{J}_{i}\left(\hat{\mathbf{r}}\right) \right)  \right],
\end{split}
\end{equation}
where the position of UAV at the epoch $t+k$ is given by \citep{Uluskan20:Noncausal}:
\begin{equation}
\begin{split}
x^\mathrm{UAV}_{m,t+k} \left( \alpha \right) &= x^\mathrm{UAV}_{m,t} + k \cdot l \cdot \mathrm{cos} \left( \alpha \right)\\[0.6em]
y^\mathrm{UAV}_{m,t+k} \left( \alpha \right) &= y^\mathrm{UAV}_{m,t} + k \cdot l \cdot \mathrm{sin} \left( \alpha \right).
\end{split}
\end{equation}

In Equation (\ref{eq:Predictive}), the predictive approach considers the FIM of all the future values along the UAV heading direction from time $t+1$ to $N$. \citet{Uluskan20:Noncausal} analyzed that the predictive approach induces UAVs to move in a linear motion, whereas the greedy approach induces a curved motion by their nature.

In \citet{Uluskan20:Noncausal}, it has been shown that the predictive approach is especially effective for UAVs with limited capability (e.g., time limitation or battery capacity) on traveling. 
The previous study showed that the predictive approach is effective in reducing localization error at the end of the search ``when the total travel length is shorter than the initial distance to the target'' \citep{Uluskan20:Noncausal}. 
We consider an emergency situation with a time constraint on search. 
Therefore, we mainly assumed in our simulation that the overall travel length of each UAV is less than the initial distance between the UAV and target.

\section{Evaluation} \label{sec:evaluation}

\begin{figure}
    \centering
    \includegraphics[width=0.8\linewidth]{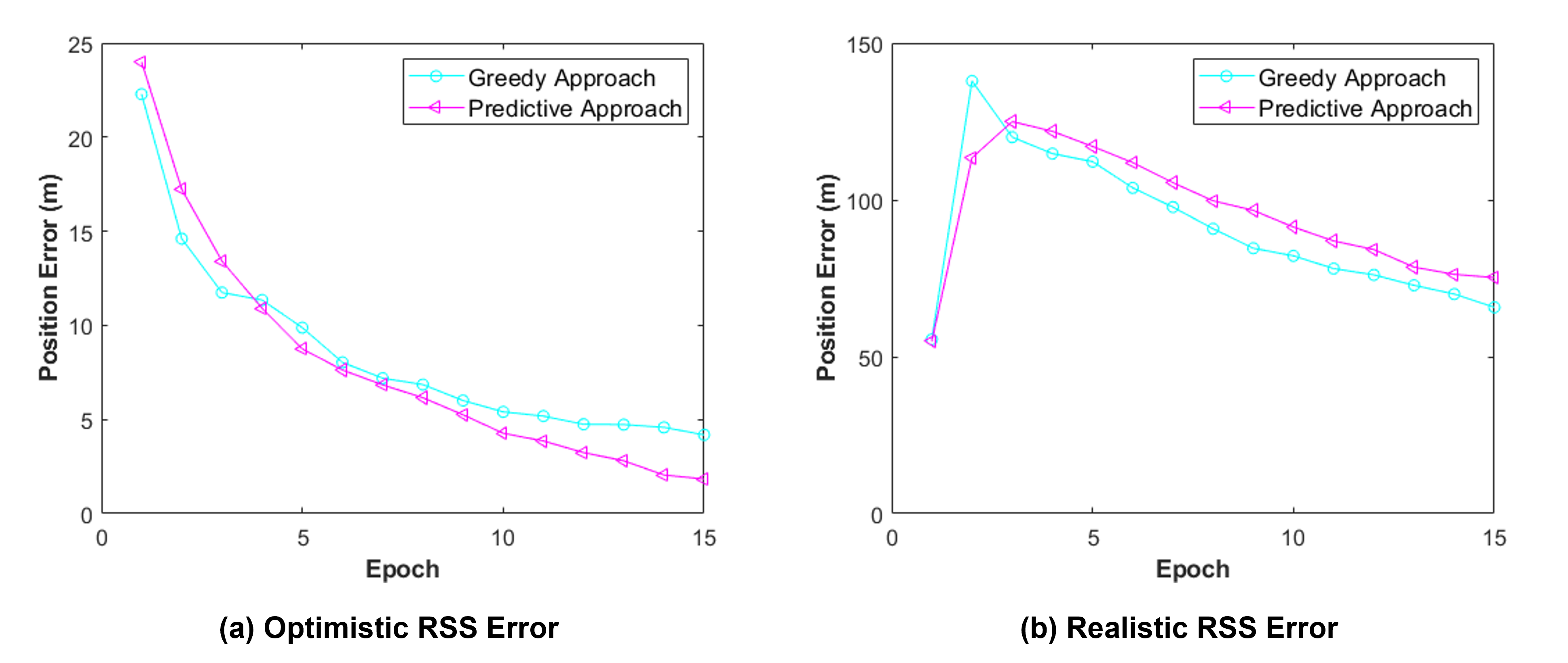}
    \caption{Target Localization Performance of the Greedy and Predictive Approaches Using a Single UAV under (a) Optimistic and (B) Realistic RSS Error Assumptions}
    \label{fig:singleUAV_result}
\end{figure}

We evaluated the target localization performance of the greedy and predictive approaches according to optimistic or realistic RSS error assumptions.
Based on the performance observations, a hybrid application of the greedy and predictive methods is proposed. 
The performance of the proposed hybrid approach is evaluated under realistic UAV geometries and RSS error assumptions.

\subsection{Single UAV Scenarios} \label{subsec:SingleUAVScenarios}

Single UAV simulations were performed with $\sigma_{\mathrm{dB}}$ set to 1) optimistic (i.e., 0.01 dB) and 2) realistic (i.e., 6 dB) values. 
Figure \ref{fig:singleUAV_result} shows the target localization performance of greedy and predictive approaches with (a) optimistic and (b) realistic RSS error assumptions. 
In both cases, the UAV started the mission at (0 m, 100 m) and the target was fixed at (0 m, 0 m). 
UAV flied 5 m per epoch (i.e., $l$ was set to 5 m). 
The total number of epochs was set to 15 for both optimistic and realistic RSS error cases.
In both cases, 100 simulations were conducted, and the root-mean-squared error (RMSE) was calculated. 
$P_0$ and $\beta$ were set to 10 dBm and 3, respectively.
The optimization problems of Equations (\ref{eq:MLE}), (\ref{eq:Greedy}), and (\ref{eq:Predictive}) were solved by the grid search-based MLE solver that we implemented. 
The distance increment for the grid search for localizing the target using Equation (\ref{eq:MLE}) was set to 1 m.
The directional angle increment for the grid search for finding the optimal flying direction of UAVs using Equations (\ref{eq:Greedy}) and (\ref{eq:Predictive}) was set to $5^{\circ}$.

\begin{figure}
    \centering
    \includegraphics[width=0.8\linewidth]{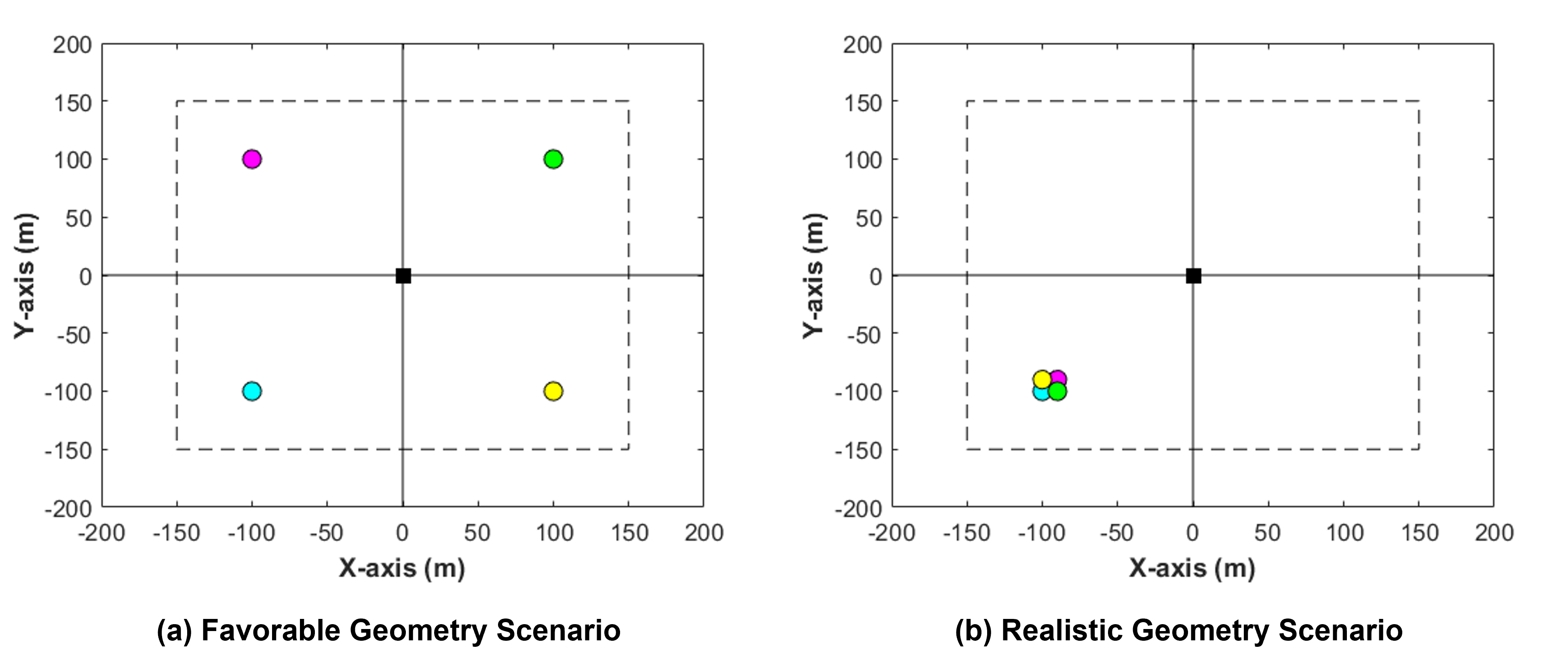}
    \caption{Initial Position of UAVs (Colored Circles) and Target (Black Square) in (a) Favorable Geometry Scenario and (B) Realistic Geometry Scenario
    }
    \label{fig:geometries}
\end{figure}

\subsection{Hybrid Application of the Greedy and Predictive Methods}

According to the simulation results in Figure \ref{fig:singleUAV_result}, the greedy approach shows better localization accuracy in the realistic RSS error case and the initial stage of the optimistic RSS error case.
However, after the localization error reduces to a certain value (e.g., approximately 10 m in the case of Figure \ref{fig:singleUAV_result}(a)), the predictive approach shows better accuracy than the greedy approach.
When the target localization error is large, the curved trajectories induced by the greedy approach are beneficial to diverging UAV geometry. 
However, after the localization error is decreased below a certain level (that is, when the UAV geometry has sufficiently diverged), the linear trajectories induced by the predictive approach become more effective in minimizing the localization error variance.
Hence, we propose a hybrid scheme of greedy and predictive approaches: using a greedy approach at the beginning of the target search and then using a predictive approach after the target localization error reduces to a certain extent. 
The time of switching from greedy to predictive can be determined adaptively. 
However, we selected the switching time manually in this preliminary study.

\subsection{Multiple UAV Scenarios} \label{sec:evaluation_multiple}

We first evaluated the greedy and predictive approaches in scenarios when the geometries between the target and UAVs are 1) favorable (i.e., all UAVs are evenly distributed around the target) and 2) realistic (i.e., all UAVs depart from the same base). 
Figure \ref{fig:geometries} shows the initial positions of the UAVs and target in two scenarios. 
The UAVs are represented by colored circles, while the target is represented by a black square. 
In the favorable geometry scenario, four UAVs are evenly deployed on a square with a side length of 200 m that surrounds the target. In the realistic geometry scenario, four UAVs start the mission at ($-100$ m, $-100$ m) and the target is fixed at (0 m, 0 m). In both two scenarios, $\sigma_{\mathrm{dB}}$ was set to 6 dB. $l$ was set to 5 m and the total number of epochs was set to 27. 
A total of 27 epochs was selected because we set ``the ability to reach the target'' of the predictive approach to 0.95. 
The ability to reach the target refers to the ratio of the total travel distance to the initial distance between the UAV and target, which is discussed in \cite{Uluskan20:Noncausal}.
In both scenarios, 100 simulations were performed. 
The settings for MLE were the same as the single UAV simulation case discussed in Section \ref{subsec:SingleUAVScenarios}.
The range for the grid search was set to 300 m $\times$ 300 m, which is shown as dashed squares in Figure \ref{fig:geometries}.

\begin{figure}
    \centering
    \includegraphics[width=0.8\linewidth]{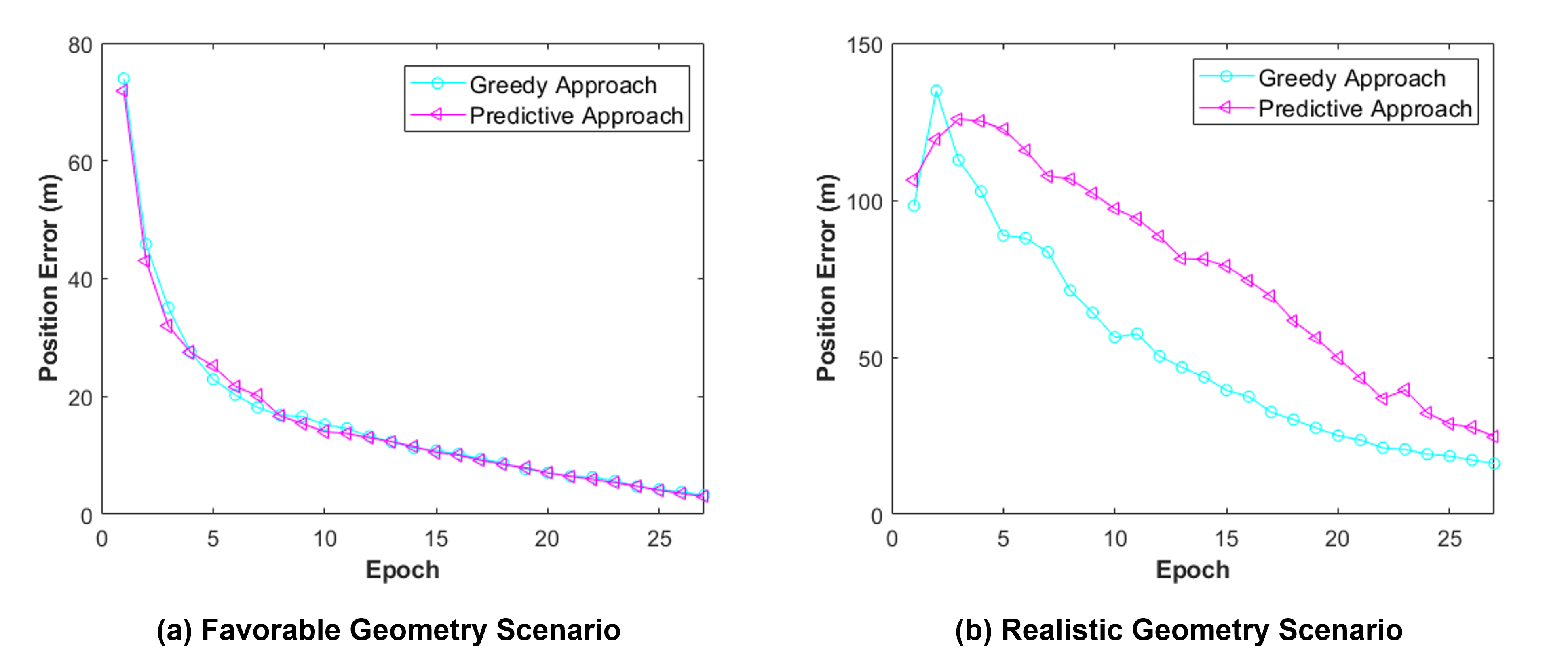}
    \caption{Target Localization Performance of the Greedy and Predictive Approaches Using Four UAVs under (a) Favorable and (B) Realistic Geometry Scenarios}
    \label{fig:multipleUAV_result}
\end{figure}

Figure \ref{fig:multipleUAV_result} shows the simulation results. 
In the favorable geometry scenario, both approaches demonstrated similar target localization performance. 
This is because both approaches induced UAVs to fly similar trajectories (i.e., along straight lines toward the target). 
On the other hand, the greedy approach was more effective in minimizing the target localization error in the realistic geometry scenario. 
While both approaches performed well in the favorable geometry scenario, a more effective approach is needed in the realistic geometry scenario. 
Hence, we applied the proposed hybrid approach to the realistic geometry scenario.

Table \ref{tab:sim_results} shows the localization performance of the proposed hybrid approach in the realistic scenario. 
The final target localization errors after 27 epochs are compared.
In the hybrid approach, the greedy approach was applied for the first 10 epochs, and then the mode was switched to the predictive approach. 
Even with this simple scheme, the hybrid approach improved the target localization performance by 30.8\% and 55.0\%, compared to the greedy-only and predictive-only approaches, respectively.

\begin{figure}
    \centering
    \includegraphics[width=0.65\linewidth]{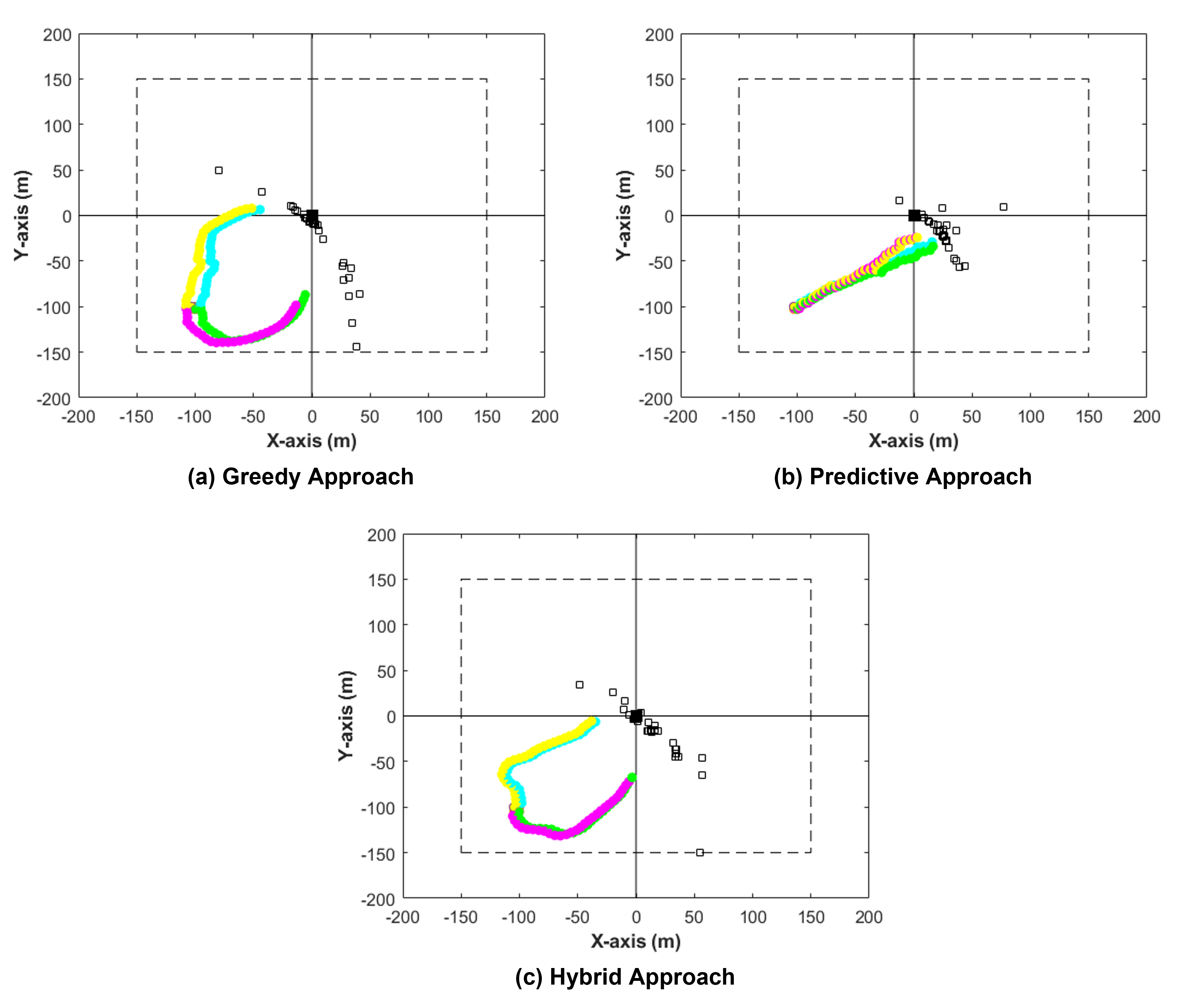}
    \caption{Example UAV Trajectories (Colored Circles) under (a) Greedy, (b) Predictive, and (c) Hybrid Approaches (The True Position of the Target is Marked as a Black Square at the Origin and the Estimated Positions of the Target are Marked as Empty Squares)}
    \label{fig:UAV_trajectories}
\end{figure}

Figure \ref{fig:UAV_trajectories} shows example UAV trajectories under the three approaches. 
The greedy approach makes UAVs follow curved trajectories toward the target at the origin, as shown in Figure \ref{fig:UAV_trajectories}(a). 
In contrast, the predictive approach in Figure \ref{fig:UAV_trajectories}(b) generates linear trajectories. 
During the initial period of search, our hybrid approach in Figure \ref{fig:UAV_trajectories}(c) yields curved trajectories as the greedy approach case. 
After the first 10 epochs, linear trajectories are generated as the predictive approach case.

\begin{table}
\small
\centering
\caption{Final Target Localization Errors of Three Approaches}\label{tab:sim_results}
\begin{center}{
\renewcommand{\arraystretch}{1.4}
 \begin{tabular}{ P{3cm} | P{2.5cm} | P{2.5cm} | P{2.5cm}}
 \hhline{====}
 \rule{0pt}{15pt} \thead{Approach} & \thead{Greedy} & \thead{Predictive} & \thead{Hybrid \\ (proposed)} \\
 \hline
 \rule{0pt}{20pt} \thead{Final Position \\ Error (m)} & 16.12 & 24.78 & 11.15\\
 \hhline{====}
\end{tabular}}
\end{center}
\end{table}

\section{Conclusion} \label{sec:conclusion}

In emergency situations, the location information of a target mobile device based on GNSS is not always available to first responders. 
Therefore, a backup system for localizing the target is necessary. 
One way to localize the target is to use UAVs in an emergency site for directly searching the target based on the received radio signals transmitted from the target. 
UAVs in an emergency should fly efficiently to find the location of the target in the shortest time possible. 
Previous research suggested two UAV trajectory optimization methods: greedy and predictive approaches. 
However, the performance of the two approaches was evaluated only under optimistic scenarios where the RSS error is small and the geometry between the UAVs and target is favorable.

Hence, we first evaluated the two approaches with single UAV simulations in an optimistic RSS error case and a realistic RSS error case where the RSS error satisfies the ITU-R recommendation. 
In the simulation results, the greedy approach showed better localization accuracies in the realistic RSS error case and the initial stage of the optimistic RSS error case. 
However, after the localization error converged to a certain value (e.g., about 10 m in our simulation), the predictive approach showed better accuracy than the greedy approach in the optimistic RSS error case.

Based on these observations, we propose a simple hybrid approach using both the greedy and predictive approaches. 
The performance of the hybrid approach was evaluated in a realistic scenario where the UAV geometry is unfavorable (i.e., four UAVs depart from the same location) and the RSS error is large.
According to the simulation results, the hybrid approach provides a better localization accuracy than the cases of greedy and predictive approaches.

\section*{ACKNOWLEDGEMENTS}

This research was supported by the Future Space Navigation \& Satellite Research Center through the National Research Foundation funded by the Ministry of Science and ICT, Republic of Korea (2022M1A3C2074404).
This research was also supported by the Unmanned Vehicles Core Technology Research and Development Program through the National Research Foundation of Korea (NRF) and Unmanned Vehicle Advanced Research Center (UVARC) funded by the Ministry of Science and ICT, Republic of Korea (2020M3C1C1A01086407).

\bibliographystyle{apalike}
\bibliography{mybibfile, IUS_publications}

\end{document}